\documentstyle[11pt,aaspp4]{article}

\lefthead{Nelson et al.}
\righthead{Dramatic Spin-Up/Spin-Down Torque Reversals}

\def\etal{et~al.}

\begin{document}

\lefthead{Nelson et al.}
\righthead{Spin-up/Spin-down Transitions in Accreting Pulsars}

\title{On the Dramatic Spin-up/Spin-down \\ Torque Reversals in Accreting Pulsars}

\author{Robert~W.~Nelson\altaffilmark{1},
Lars~Bildsten\altaffilmark{2},Deepto~Chakrabarty\altaffilmark{3},
\\ Mark H. Finger\altaffilmark{4},Danny~T. Koh\altaffilmark{5},
Thomas~A.~Prince\altaffilmark{5},Bradley~C.~Rubin\altaffilmark{4,6},
\\D. Mathew Scott\altaffilmark{4}, Brian~A.~Vaughan\altaffilmark{5} 
and Robert~B.~Wilson\altaffilmark{4}}


\altaffiltext{1}{Theoretical Astrophysics, California Institute of
Technology, Pasadena, CA 91125} 
\altaffiltext{2}{Department of Physics and Department
  of Astronomy, University of California, Berkeley, CA 94720}
\altaffiltext{3}{Center for Space Research, Massachusetts
Institute of Technology, Cambridge, MA 02139}
\altaffiltext{4}{Space Science Laboratory, NASA/Marshall Space
  Flight Center, Huntsville, AL 35812}
\altaffiltext{5}{Space Radiation Laboratory, California Institute
of Technology, Pasadena, CA 91125}
\altaffiltext{6}{
Cosmic Radiation Lab, RIKEN Institute, 
Wako-shi, Saitama 351-01, Japan}

\begin{abstract}
Dramatic torque reversals between spin up and spin down
have been observed in half of the persistent X-ray pulsars 
monitored by the BATSE all-sky monitor on CGRO.
Theoretical models developed to explain early pulsar timing 
data can explain spin down torques via a disk-magnetosphere interaction 
if the star nearly corotates with the inner accretion disk.
To produce the observed BATSE torque reversals, however, these
equilibrium models require the disk to alternate
between two mass accretion rates, with $\dot M_{\pm}$
producing accretion torques of similar magnitude, but always of opposite sign.
Moreover, in at least one pulsar (GX 1+4) 
undergoing secular spin down the neutron star spins down faster 
during brief ($\sim 20$ day) hard X-ray 
flares -- this is opposite the correlation expected from standard 
theory, assuming BATSE pulsed flux increases with 
mass accretion rate. The $10$ day to 10 yr intervals 
between torque reversals in these systems are much longer than 
any characteristic magnetic or 
viscous time scale near the inner disk boundary and are more
suggestive of a global disk phenomenon.

We discuss possible explanations of the observed torque behavior.
Despite the preferred sense of rotation defined by the binary orbit, 
the BATSE observations are surprisingly 
consistent with an earlier suggestion by Makishima \etal (1988) for GX~1+4:
the disks in these systems somehow alternate between 
episodes of prograde and retrograde rotation. We are unaware 
of any mechanism that could produce a stable retrograde disk 
in a binary undergoing Roche-lobe overflow, but such flip-flop behavior 
does occur in numerical simulations of wind-fed systems. 
One possibility is that the disks in some of these binaries 
are fed by an X-ray excited wind. 
\end{abstract}


\section{Spin Up and Spin Down in Accretion-Powered Pulsars}

The spin evolution of an accreting magnetic star, 
an X-ray pulsar, magnetic CV, or T Tauri star, 
is thought to be regulated by torques acting between 
the accretion disk and the stellar magnetosphere
(Rappaport \& Joss 1977; Warner 1990; Konigl 1991).
Because of the small neutron star moment of inertia, however, 
only the X-ray pulsars undergo accretion-induced 
changes in rotation frequency large enough 
to measured on short time scales ($\sim$ days).
They are thus ideal laboratories for studying the 
dynamical interaction between a magnetic star and 
its accretion disk.

Accreting X-ray pulsars are rotating, highly magnetized 
($B \sim 10^{12}$\,G)
neutron stars that accrete material from a stellar companion, either
from a stellar wind, or by Roche-lobe overflow mediated by an
accretion disk.  Disks may also form in wind-fed systems if the
captured material has sufficient angular momentum to circularize
before reaching the neutron star magnetosphere (see King 1995).
The strong magnetic field disrupts the disk and forces the accreting
plasma to corotate with the star at a radius
where magnetic and fluid stresses roughly balance,  
$r_m \sim \mu^{4/7} \dot M^{-2/7} (GM_x)^{-1/7} \sim 10^{8-9}$ cm, 
where $\dot M$ is the mass accretion rate, 
$\mu$ is the magnetic dipole moment, and $M_x$ is the 
mass of the neutron star. Although the coupling between the 
disk and magnetosphere is complicated, and may depend on the geometry 
and relative orientation of the magnetic field (Wang 1997), in the simplest picture of 
accretion torque (Pringle \& Rees 1972; Rappaport \& Joss 1977)
one assumes that the specific angular momentum of material captured from the
inner accretion disk is somehow transported onto the star with 
the accreting matter. For a Keplerian disk, the pulsar will experience a spin-up torque 
\begin{equation}
N =  \dot M\sqrt{GM r_m}=2\pi I \dot \nu, 
\label{eq:torque}
\end{equation}
where $\dot \nu$ is the pulsar spin frequency, and 
$I \sim 10^{45}{\,\rm gm\,cm^2}$ is the neutron-star 
moment of inertia.
Early observations indicated that, on average, 
most accreting pulsars were spinning up on a time scale
$t_{su}= {\nu/\dot \nu} \sim  10^{4}$ yrs
consistent with equation (\ref{eq:torque}). 
This was strong evidence that X-ray pulsars must be compact stars
with large magnetospheric radii (Rappaport \& Joss 1977). 

This simple spin-up picture had to be modified when 
two well-studied pulsars, Her X-1 and Cen X-3, were
found to be spinning up much more slowly than predicted by
equation (\ref{eq:torque}). Furthermore, these pulsars sometimes underwent 
short episodes of spin-down (Elsner \& Lamb 1977; Ghosh \& Lamb
1979). How could a star capturing material from a disk with the
same sense of rotation actually lose angular momentum while continuing
to accrete? 

To explain this behavior Ghosh and Lamb (1979; hereafter GL) 
argued that the spin-up accretion torque must decrease -- 
and eventually become negative -- when the stellar rotation frequency 
approaches the Keplerian orbital frequency of the inner accretion disk, 
$\Omega_* \simeq \Omega_K(r_m) =(GM/r_m^3)^{1/2}$. 
Since most X-ray pulsars are in binaries much older than the
pulsar spin-up time scale, GL argued they
should have reached this near-equilibrium state.
In this situation, magnetic field lines which thread the disk 
beyond the corotation radius (where the disk rotates more 
slowly than the star) are swept back in a trailing spiral 
and transport angular momentum outward. 
Stars close to equilibrium will spin up much more slowly than 
predicted by equation (\ref{eq:torque}) and 
can even spin down while continuing to accrete.

\placefigure{fig:fig1}

GL wrote their torque as a modified form of equation
(\ref{eq:torque}), 
\begin{equation}
N_{GL} = n(\omega)\dot M\sqrt{GM r_m} 
\label{eq:gltorque}
\end{equation}
where $n(\omega)$ is a dimensionless function of the ``fastness
parameter'', $\omega = {\Omega_*/\Omega_K(r_m)} \propto 
\dot M^{-3/7}\Omega_* B_*^{6/7}$.
For most observations $\Omega_*$ can be taken as constant,
so that, in their theory, observed torque fluctuations mainly reflect 
the dependence of $N_{GL}$ on $\dot M$
Although several functional form has been 
suggested (e.g. Campbell 1987; Wang
1987, 1995, 1997), for our discussion it is only important that
$N_{GL}$ is a smooth and monotonically increasing function of 
$\dot M$ that crosses zero at some 
$\dot M_{crit}$ corresponding to a critical fastness parameter 
$\omega_{crit} \la 1$. An approximate version of
$N_{GL}(\dot M)$ with $\omega_{crit}=0.8$ is shown in Figure 1.
In particular, the spin-up torque becomes negative
at low accretion rates; the magnetospheric radius 
$r_m \propto \dot M^{-2/7}$ moves outwards,
close enough to the corotation radius $r_{co}=(GM/\Omega_*^2)^{1/3}$ 
that the negative magnetic torques become large. 
Note, however, that sudden changes in accretion torque require sudden 
changes in $\dot M$.


\section{Observations of Torque Reversals with BATSE}

Prior to 1991, the spin periods of accreting pulsars were typically
measured only once or twice per year by pointed X-ray telescopes
(Nagase 1989 and references therein).
Consequently, published measurements of accretion torque were usually 
long-term averages. 
Since the launch of the Compton Gamma Ray Observatory in April 1991, 
however, the BATSE instrument has compiled the first continuous long-term 
history of pulse frequencies  for the majority of persistent X-ray pulsars 
(Bildsten et al. 1997), increasing the sampling of pulse periods by 
more than a factor of 100.

\placefigure{fig:fig2}

The frequency history of the 4.8\,s pulsar Cen X-3 shown in 
Figure \ref{fig:fig2} is one example where BATSE
observations reveal a strikingly different picture of pulsar spin
behavior than previously understood. Prior to BATSE, the long-term
frequency history had been described as secular spin-up at $\dot \nu =
8\times10^{-13}$\,Hz\,s$^{-1}$ -- a factor of $\sim$5 slower than
predicted by eq.~\ref{eq:torque} -- superposed with wavy fluctuations
and short episodes of spin down.  This behavior was interpreted 
as evidence that Cen X-3 was rotating near its equilibrium 
spin period with a significantly reduced torque (Elsner \& Lamb 1977).
In contrast, the more frequently sampled BATSE 
data show that Cen X-3 actually exhibits frequent transitions between 
states of steady spin up and spin down, with a short-term torque 
magnitude consistent with equation ~\ref{eq:torque}.
The bimodal torque behavior
has been confirmed quantitatively (Finger, Wilson \& Fishman 1994).  The
pulsar is nearly always in one of two possible torque states ($\dot
\nu \simeq -4 \times 10^{-12} \rm{Hz\,s^{-1}}$ or $\dot \nu \simeq 7
\times 10^{-12} \rm{Hz\,s^{-1}}$) and remains in one state for $\sim
10-100$ days before switching to the other.
Transitions between spin up and spin down occur on a time scale more
rapid than BATSE can resolve, $\la$10\,d.  
It is now evident that the reduced long-term spin-up torque inferred 
from the sparse pre-BATSE data is a consequence of these 
frequent transitions between spin up and spin down
(Prince et al. 1994). 

Interestingly, torque transitions like those seen in Cen X-3 appear 
to be common:  at least 4 out of the 8 persistent pulsars 
observed by BATSE show torque reversals between steady spin-up 
and steady spin-down.  Of the remaining four, three
(GX 301--2, Vela X-1 and 4U 1538-52) are wind-fed pulsars while
Her X-1 is sampled infrequently at 35 day intervals so that we cannot
measure its torque on short time scales. 
One of the most dramatic
torque transitions took place in the $7.6$s pulsar 4U 1626-67 around
1991 (Chakrabarty et al. 1997a).  After two decades of the smoothest
spin up observed in any accreting pulsar, BATSE found that 4U1626-67
was smoothly spinning down.  Most surprisingly, the spin down torque
is nearly equal in magnitude, but opposite in sign, to the spin up 
torque. A similar
transition to spin down was observed in the 120\,s pulsar GX 1+4 in
1988 (Makishima et al. 1988) after more than a decade of steady spin
up.  Again, the spin down rate is close in magnitude to the spin up
rate.  Finally, the 38\,s disk-fed pulsar OAO~1657--415 shows torque
episodes with strength and duration
very close to those seen in Cen X-3 (Chakrabarty et al. 1993).

\section{Accretion from Retrograde Disks?}

To explain the BATSE observations, the near-equilibrium models
described above require the 
disks in systems like Cen X-3 to somehow undergo repeated step-function-like 
changes in the mass accretion rate (Figure 1). Moreover, these transitions 
in $\dot M$ must always produce torques of
opposite signs, but with similar magnitudes. Indeed, the hypothesized
change in accretion rate in 4U 1626-67 
must result in a spin-down torque within $15\%$ 
of its previous spin-up rate. On the other hand, if the cycle of 
transitions reflects physics occurring at the boundary between the 
disk and the magnetosphere, then one would expect the characteristic 
time between transitions to range between the dynamical time scale,
$t_d \sim \Omega^{-1}_K \sim 1$s, and the inner disk viscous time scale,
$t_v \sim  R/H (\alpha \Omega_K)^{-1} \sim 10^3$ s, where $H$ is the
disk thickness. Yet systems like GX 1+4 and 4U 1626-67 were stable 
for years before reversing their torques. 

When the transition from spin up to spin down was first detected in
GX 1+4, Makishima et al. (1988) instead suggested that the previous disk had 
dissipated and a new disk had formed with a reversed sense of
rotation.  Instead of undergoing mass transfer by Roche-lobe overflow,
they argued that GX 1+4 is accreting from a dense, subsonic 
wind from its M giant companion; transient formation of alternating 
prograde and retrograde disks are known to occur in numerical simulations of 
wind-fed systems (Fryxell \& Taam 1988). The formation of a retrograde disk 
would also explain the similar torque magnitudes in both states, and 
obviates the need for an ultrastrong magnetic dipole field, 
$B \simeq 10^{14}$ G, if this slow pulsar ($P_p \sim 120$ s) is to 
corotate with the inner accretion disk, $\Omega_*=\Omega_K(r_m)$.

Chakrabarty et al. (1997b) have recently found
surprising evidence, however, which supports the presence of a retrograde
accretion disk in GX 1+4. While in an extended
spin-down state, GX 1+4 spins down more rapidly during short-term 
20-50 keV ($\sim 20$ day) flares observed with BATSE. That is, 
the torque is {\it anticorrelated} with the observed flux.  
This is opposite the prediction of standard 
spin-down theory (see Fig 1.). At higher accretion
rates the magnetosphere should move inward away from corotation,
reducing the magnetic spin-down torques while increasing the material
spin-up torques.  On the other hand, if GX 1+4 is accreting from a 
retrograde disk, one expects the spin-down rate to increase with luminosity.
In that case the material really does carry negative angular momentum
relative to the neutron star rotation.


Could the torque transitions seen in other X-ray pulsars 
also be due to alternating episodes of prograde and 
retrograde rotation? This hypothesis would naturally 
explain several puzzling aspects of the BATSE observations. 
If the disks in these systems are somehow produced with both senses 
of rotation one expects the observed two-state torque behavior 
-- the material can only circulate one way or the other. 
Moreover, transitions between torque
states of the same sign will never occur, and for comparable 
mass accretion rates the torques should have comparable magnitudes; 
in the simplest picture, $N \simeq \pm \dot M \sqrt{GMr_m}$.
The time intervals between torque reversals are also consistent with 
the global disk viscous times in these systems 
$t_v \sim R^2_D/\nu \sim $ days - yrs for a fully ionized 
Shakura-Sunyaev disk with $\alpha \sim 0.01-0.1$.
One can imagine a cycle where a disk form, accretes all of its
material, and then a new disk forms with the opposite sense of rotation.

If these binaries undergo mass transfer by standard Roche-lobe 
overflow we know of no mechanism that could produce a 
stable retrograde disk. The specific angular momentum initially 
carried by the accretion stream $l \sim d^2 \Omega_{orb}$ 
($d$ is the distance from the neutron star to the first Lagrange point,
and $\Omega_{orb}$ is the orbital frequency)
is comparable to the specific orbital angular momentum of 
the companion star and should circularize in the prograde sense well before 
reaching the pulsar magnetosphere (Lubow \& Shu 1975).
However, transient disks with alternating sense of 
rotation are known to form in numerical simulations of 
non-axisymmetric wind-fed accretion (Fryxell and Taam 1988; Ruffert
1997). It may be possible that the disks in these  systems 
are fed from a stellar wind, rather than from Roche-lobe overflow
as is commonly assumed. Davidsen, Malina \& Bowyer (1977).
suggested that GX ~1+4 accretes from a slow dense wind  from its 
M-giant companion. The massive OB-type companion of Cen X-3
is close to Roche-lobe filling (van Paradijs \etal 1983, 
Chakrabarty et al. 1993), but should also have a strong 
stellar wind (Day \& Stevens 1993), as should the massive companion
of  OAO~1657--415. 

The ultra-compact binary 4U 1626-67 is a more
challenging case.  The small upper limit on its mass function 
$f_x(M) < 10^{-6} M_\odot$ and probable 42-min orbit is usually 
interpreted as indicating a very
low-mass hydrogen-depleted main sequence star or degenerate He dwarf 
undergoing Roche-lobe overflow
(Verbunt, Wijers \& Burm 1990; Chakrabarty et al. 1997a). 
Based on {\sl ASCA} detection of neon 
emission lines near 1 keV, Angelini et al. (1995) suggested the
companion star may be a helium burning star with a very strong mass outflow.
However, this scenario requires a very small and improbable orbital 
inclination ($i < 1^\circ.6$ for $M_c = 0.6 M_\odot$) and high mass outflow.
Alternatively, the X-ray flux from the pulsar itself may induce 
a self-excited wind in this system (Basko \etal 1977; Tavani \& London 1993). 
Indeed, for a distance $D=5 D_5$ kpc, the incident X-ray flux at 
the companion surface is a factor $\sim 50 D_5^2$ larger than 
the flux incident on the companion of  Her X-1, the system for which the 
X-ray excited mass loss models were invented. 
Assuming the companion is an $M=0.08 M_\odot$ main-sequence star, 
we estimate an induced mass outflow 
$\dot M_w \sim 2 \times 10^{16} D_5^2 \rm{\,gm\,s^{-1}}$ (Basko \etal 1977).
This is about 5 times the inferred mass accretion rate;
efficient capture of the wind is expected since the 
more massive neutron star dominates the binary potential well.

\section{Discussion}

We wish to make two points in this paper. First, the highly-sampled 
BATSE pulsar timing data is difficult to reconcile with standard
explanations of spin-down accretion torques in near-equilibrium 
rotators, $\Omega_*=\Omega_K(r_m)$.
According to these theories, the switching between spin up 
and spin down now seen in the majority of persistent BATSE pulsars 
would require repeated transitions 
between two mass accretion rates, with $\dot M_\pm$ 
producing torques of comparable
magnitude, but always of opposite sign. Moreover, in one pulsar, the 
observed anticorrelation between torque and 20-60 keV pulsed
luminosity is opposite the predicted effect. Some bistable torque
mechanism with a switching time scale much longer than any natural 
time at the inner disk boundary must be at work. This time scale 
is more consistent with a global disk phenomenon. 

Recently, Yi, Wheeler \& Vishniac (1997) have suggested that the 
observed torque reversals may be due to a transition from a standard 
Keplerian disk rotation law to a sub-Keplerian advection-dominated
flow (ADF); they write the new rotation law
$\Omega^\prime(r)=A\Omega_K(r)$ with $A=0.2$ and  assume the star 
is initially spinning near equilibrium.
The sudden transition to ADF decreases the corotation radius, 
$r^\prime_{co} = A^{2/3}r_{co}$, while the fastness parameter increases 
$\omega^\prime = \omega/A \propto \dot M^{-3/7}B_*^{6/7}/A$.
Assuming a GL-type torque (eq. \ref{eq:gltorque}) in both 
states, it is not surprising that Yi \etal (1997) find
acceptable fits to the observed transitions: for a given A, 
one can always adjust $B_*$ and $\dot M$ to yield the observed 
torques before and after the transition.  Their fitted
parameters, however, do not agree with observational constraints. 
$\dot M$ must be near the critical accretion rates required by ADF, 
$\dot M_{crit} \sim 0.1 \alpha^2 \dot M_{Edd}
\sim 10^{15}-10^{16}\rm{gm\,s^{-1}}$ (Narayan \& Yi 1995), yet 
Cen X-3, GX 1+4 and OAO 1657-415 are accreting at much 
higher rates  $\dot M \sim 10^{17}-10^{18} \rm{\, gm\,s^{-1}}$ (Nagase 1989; 
Chakrabarty \etal 1993, 1997b). Moreover, the fitted magnetic 
field  strengths are at least an order of magnitude smaller than
required for these stars to be near spin equilibrium.  
Nevertheless, this scenario is attractive because it is consistent
with the transition time scales discussed above -- the ADF transition itself 
will occur on a fast thermal time $t_{th} \sim (\alpha \Omega_K)^{-1} 
\sim 10^3$ s, while 
the interval between torque transitions is set by slow changes 
in the mass transfer rate occurring on a global disk viscous time scale.

Our second point is admittedly more speculative.
Based on the suggestion by Makishima \etal (1988) that 
a reversed disk formed in the binary system containing GX~1+4,
the BATSE observations themselves are 
consistent with accretion from disks having alternating 
senses of rotation. This interpretation naturally explains 
the bimodal nature of the torques, the comparable torque 
magnitudes, the torque-luminosity 
anticorrelation seen in GX 1+4, and the time scales between 
torque transitions -- without requiring 
these systems to be near spin
equilibrium. Although we are unaware of any physical 
mechanism that could produce a reversed disk in a system 
undergoing Roche-lobe overflow, we suggest that these binaries 
may be accreting from from a stellar wind, possibly 
self-excited by the X-rays coming from the accreting neutron star. 


It may be possible to test the reversed disk hypothesis with 
further observations. As in GX 1+4, the torque in any system 
undergoing secular spin-down should spin down faster with increased 
luminosity. Observations with the X-ray Timing explorer may
be able to determine the correlation between torque and 
luminosity in Cen X-3 while it is in a spin-down state. 
We estimate that a two day pointing of Cen X-3 with XTE
could make of order 5 torque measurements sensitive
to 10\% variations at the $3\sigma$ level. If the accretion 
disk is fed from an X-ray excited wind with a 
flip-flop type instability, changes in column density 
inferred from low-energy absorption may also reveal 
transitions in the structure of the accretion flow
(Day \& Stevens 1993).



This work was funded in part by NASA grants 
NAG 5-3119, NAG 5-3109, NAG 5-1458, NAG 5-3293, NAGW-4517 and NGT-51184
and the Alfred P. Sloan Foundation.

\clearpage

\figcaption[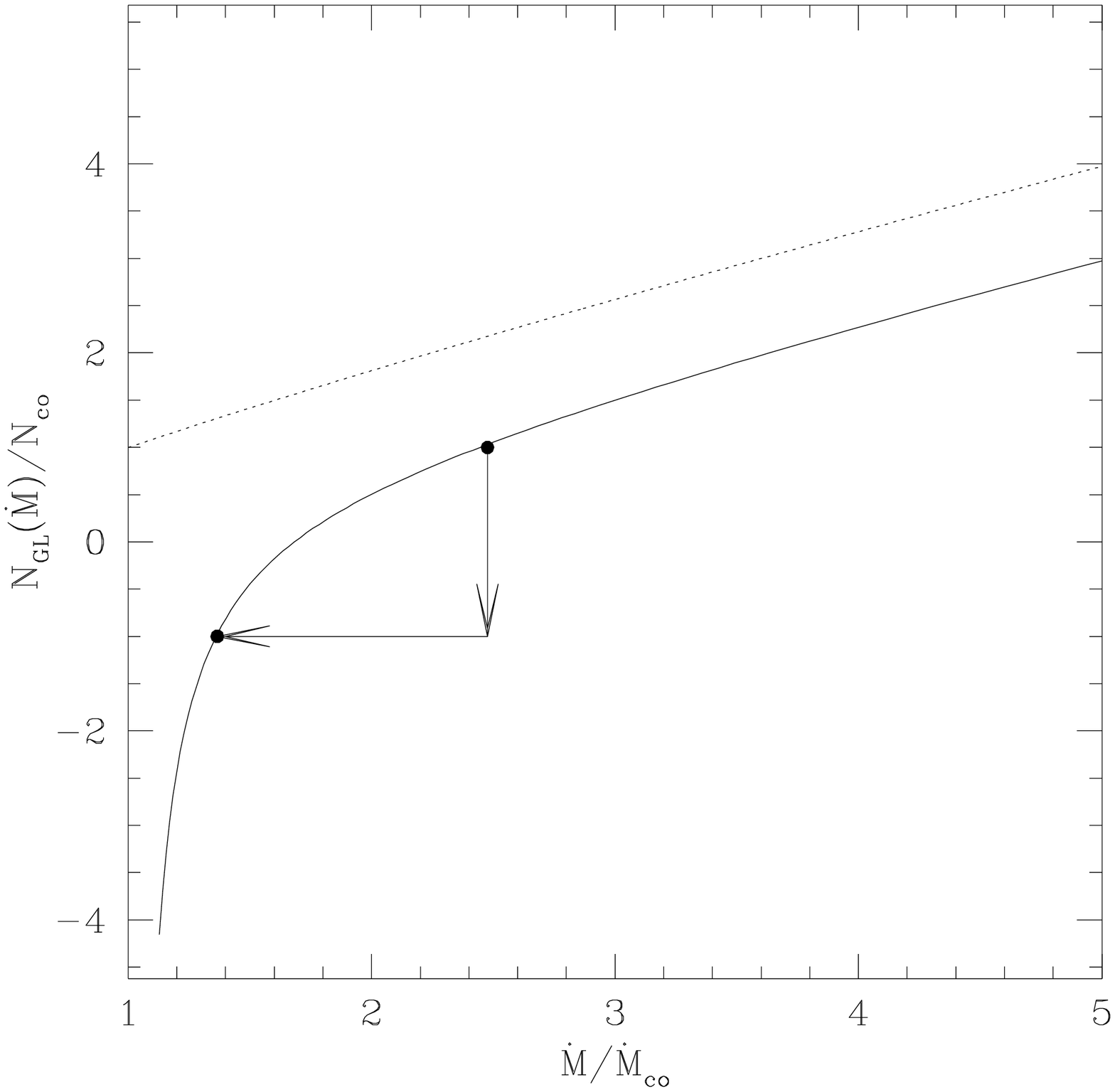]{
The Ghosh-Lamb accretion torque is a smooth function of mass accretion
rate which passes through zero and becomes negative as $\dot M$ 
decreases and the magnetosphere approaches the corotation.
Here $N_{co}=\dot M_{co}(G M r_{co})^{1/2}$
and $r_m(\dot M_{co})=r_{co}=(GM/\Omega^2_*)^{1/3}$.  
Sudden transitions between spin up and 
spin down -- as occurs repeatedly for Cen X-3 (Figure 2) -- require 
step-function-like changes in $\dot M$.
The dotted line corresponds to the simple spin-up torque model
$N=\dot M (GMr_m)^{1/2}$.
\label{fig:fig1} 
}

\figcaption[fig2.ps]{
The long term spin frequency of Cen X-3
The torque appears to be  bimodal, alternating between spin-up and 
spin-down, with an instantaneous torque magnitude much larger than 
its average value. 
\label{fig:fig2}
}

\clearpage

\plotone{fig1.eps}

\clearpage

\plotone{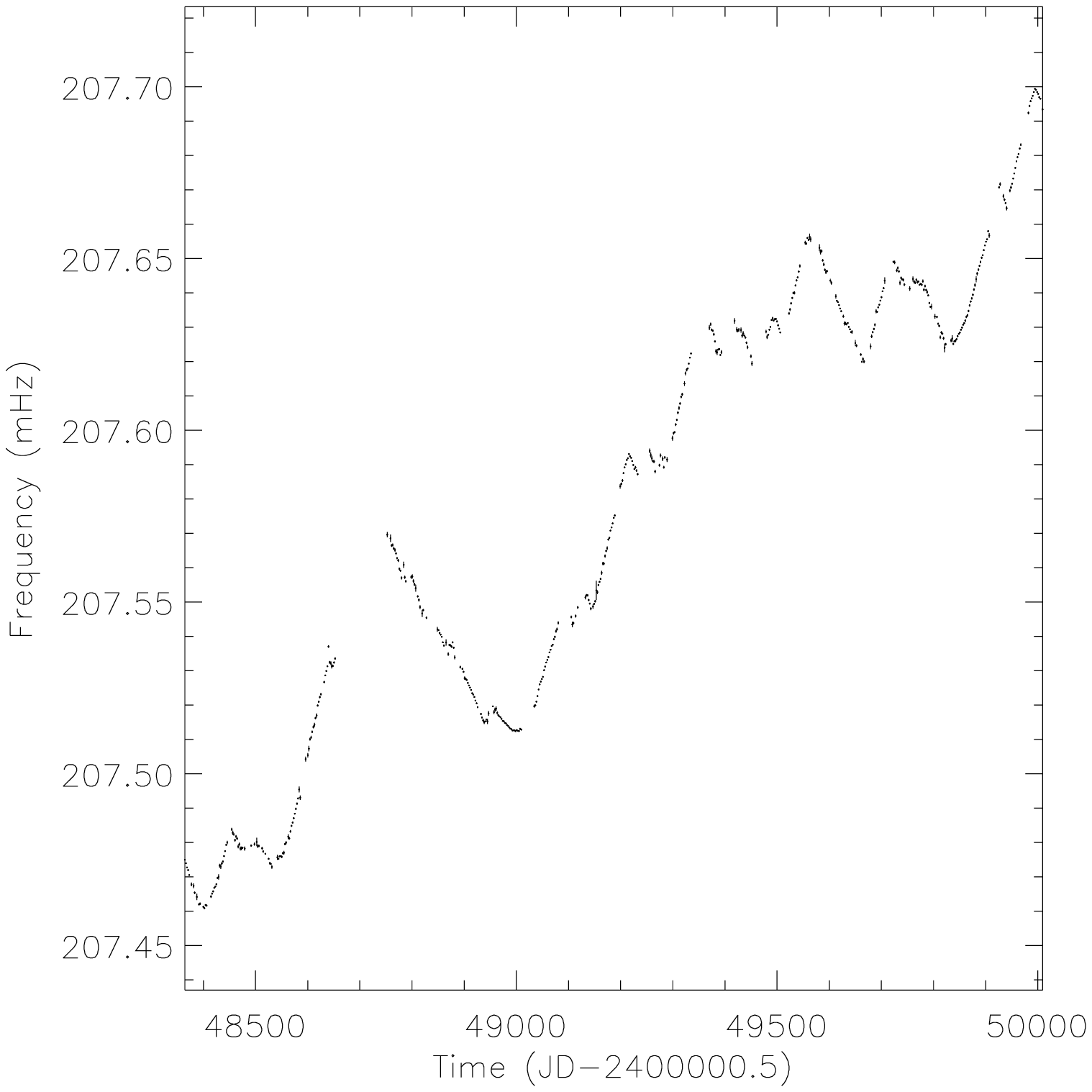}

\end{document}